\documentclass[]{aiaa-tc}

 \title{Numerical Investigation of Second Mode Attenuation over Carbon/Carbon Surfaces on a Sharp Slender Cone}

 \author{
  Victor C. B. Sousa%
    \thanks{Graduate Research Assistant, vsousa@purdue.edu}
  , Danish Patel%
    \thanks{Graduate Research Assistant, AIAA Student Member, patel472@purdue.edu.}
  , J.-B. Chapelier%
    \thanks{Post Doctoral Fellow, AIAA Member, jchapeli@purdue.edu.}  
   ,  Alexander Wagner%
    \thanks{Research Scientist, Spacecraft Department, Alexander.Wagner@dlr.de}
    \ and Carlo Scalo%
  \thanks{Assistant Professor, AIAA Member, scalo@purdue.edu.}
\\
  {\normalsize\itshape
   School of Mechanical Engineering, Purdue University, West Lafayette, IN, 47906, USA}\\
  {\normalsize\itshape
   German Aerospace Center (DLR), Institute of Aerodynamics and Flow Technology, G\"ottingen, Germany}\\
   }

 \AIAApapernumber{YEAR-NUMBER}
 \AIAAconference{Conference Name, Date, and Location}
 \AIAAcopyright{\AIAAcopyrightD{YEAR}}

\usepackage{lipsum}
\usepackage{caption}
\usepackage{subcaption}
\usepackage{amsmath}
\usepackage{wrapfig}
\usepackage{color}
\usepackage{graphicx}

\graphicspath{{./figures/}}

\begin{document}

\maketitle

\begin{abstract}
 
We have carried out axisymmetric numerical simulations of a spatially developing hypersonic boundary layer over a sharp 7$^{\circ{}}$-half-angle cone at $M_\infty=7.5$ inspired by the experimental investigations by Wagner (2015). 
Simulations are first performed with impermeable (or solid) walls with a one-time broadband pulse excitation applied upstream to determine the most convectively-amplified frequencies resulting in the range 260kHz -- 400kHz, consistent with experimental observations of second-mode instability waves. Subsequently, we introduce harmonic disturbances via continuous periodic suction and blowing at 270kHz and 350kHz. For each of these forcing frequencies complex impedance boundary conditions (IBC), modeling the acoustic response of two different carbon/carbon (C/C) ultrasonically absorptive porous surfaces, are applied at the wall. The IBCs are derived as an output of a pore-scale aeroacoustic analysis -- the inverse Helmholtz Solver (iHS) -- which is able to return the broadband real and imaginary components of the surface-averaged impedance. The introduction of the IBCs in all cases leads to a significant attenuation of the harmonically-forced second-mode wave. In particular, we observe a higher attenuation rate of the introduced waves with frequency of 350kHz in comparison with 270kHz, and, along with the iHS impedance results, we establish that the C/C surfaces absorb acoustic energy more effectively at higher frequencies.
\end{abstract}




\section{Introduction}

The design of hypersonic vehicles is constrained by the considerable heat transfer and shear stress in the boundary layer.
Although these constraints are already important in a laminar regime, they are significantly enhanced by the transition to turbulence.
Reed et al. \cite{Reed_1997} highlighted the differences in fully turbulent and fully laminar trajectories for a hypersonic flight vehicle, reporting that the vehicle would experience a heat flux approximately 5 times higher under turbulent conditions and, therefore, would require at least twice the weight in thermal protection systems (TPS).
Recent studies in the context of the national Aerospace Plane Project led by the Defense Science Board committee \cite{united1988report} have also reported that the performance of such vehicles are substantially affected by transition, reporting a three-order-of-magnitude difference in the cargo-to-weight ratio between fully laminar and fully turbulent conditions.
These studies emphasize the importance of laminar-turbulent transition delay for sustainability of high speed travel.

Hypersonic vehicles that fly with small angles of attack and with a high lift-to-drag ratio tend to have predominantly 2D geometries and slender shapes.
This configuration, in addition to the need to cool the walls in hypersonic flights, makes the second mode the main mechanism driving transition to turbulence \cite{mack1969boundary}.
The second mode was discovered by Mack \cite{mack1965computation} and consists of an acoustic wave trapped in the boundary layer.
His studies on the evolution of these and higher modes \cite{mack1990inviscid} concluded that any attempt to increase the transitional Reynolds number in high speed flows would require the stabilization of the second mode.

Fedorov et al.\cite{fedorov2001stabilization} proved through Linear Stability Theory (LST) the capability of ultrasonic absorbing coatings (UACs) to mitigate the second mode, confirmed by experiments that followed.
The first experiments to observe the predicted behavior was performed by Fedorov et al. \cite{Fedorov2001}.
They constructed a cone with a half angle of 5 degrees and divided it in two different surfaces, one smooth and the other perforated with regular micro holes.
They reported that the porous surface was capable of doubling the transitional Reynolds number in comparison with the smooth surface in experiments at Mach 5.  
%
%

The development of UACs continued and the next step was to combine the absorption and TPS characteristics in one solution.
Knowing that an irregular structure is typical for TPS materials used for hypersonic vehicles, Fedorov et al. \cite{fedorov2003stabilization,fedorov2006stability} conducted experiments in a Mach 6 wind tunnel first over a porous surface with regular microstructure, and then over an irregularly structured felt-metal surface.
Both geometries were shown to stabilize the second mode while the first mode gained importance.
%
%
Inspired by these results, Wagner et al. \cite{Wagner2013a, Wagner:2014PhD} pioneered the use of carbon-carbon (C/C), an intermediate state of C/C-SiC already employed on hypersonic vehicles\cite{TurnerHJST_AIAA_2006,WeihsLT_IAC_2008}, to control second-mode waves.
Experiments at Mach 7.5 have been conducted, and the stabilization of the second mode as well as an increase in the laminar portion of the boundary layer over the porous surface have been observed.

%
%

Previous high-fidelity numerical studies of attenuating, canceling, or reinforcing second-mode instabilities in high-speed flow over porous walls have been limited to studying the effects of uniformly spaced and geometrically regular porosity in temporally developing boundary layers \cite{bres2008stability, fedorov2011instability, bres2013second}.
The potential of ultrasonically absorptive surfaces is yet to be fully explored.
The current work plans on building upon these results by conducting a numerical study of the second mode spatial acoustic interaction with a C/C based porosity model in a hypersonic boundary layer over a sharp cone.
A simulation tool capable of reproducing the flow conditions, the transition mechanisms and the impact of a porous surface on the boundary layer stability is introduced.This new tool is designed to support the design of future randomly structured materials for transition control on hypersonic vehicles. 
%
%


\section{Physical Model and Computational Approach}

The numerical study is based on tests conducted by Wagner \emph{et al.} \cite{Wagner:2014PhD} in the DLR High Enthalpy Shock Tunnel G\"ottingen (HEG).
In the present study, the test conditions of the run with $Re_m = 4.06 \cdot 10^6$ were chosen as free stream conditions, see table \ref{tab:HEGtestconditions}.
This flow condition is applied in a simulation of a spatially developing boundary layer over a sharp cone with a half angle of 7 degrees to showcase the second-mode attenuation modeling capability of impedance boundary conditions (IBC).
 \vspace{-0.3cm}
\begin{table}
\centering
\caption{Flow parameters from Wagner's experiments\cite{Wagner:2014PhD}}
\label{tab:HEGtestconditions}

\begin{tabular}{lccc}
\hline \hline
 $Re_m$ $\Rightarrow$ & $1.46 \cdot 10^6$  & $4.06 \cdot 10^6$ & $9.8 \cdot 10^6$ \\ 
\hline                               


$M_\infty \ [-]$ & 7.3 & 7.4 & 7.4\\
$p_\infty \ [Pa]$ & 789 & 2129 & 5174\\
$T_\infty \ [K]$ & 267 & 268 & 265\\

$\rho_\infty \ [g/m^3]$ & 10.2 & 27.6 & 67.8\\
$u_\infty \ [m/s]$ & 2409 & 2422 & 2419\\

\hline \hline
\end{tabular}
\end{table}

A combination of robust low-order precursor simulations and high-order structured simulations are considered.
The former is necessary to capture the base flow around the cone geometry and it is not designed to capture transition dynamics in the boundary layer, but, rather, used to assign the correct initial conditions to the high-order structured simulations.
The precursor simulation is carried out using the unstructured code sd3dvisp, the output of which is used to provide inlet boundary conditions to CFDSU.
More details on these two codes are given in section III.

\begin{figure}[htbp]
\begin{center}
\includegraphics[width =.9\columnwidth]{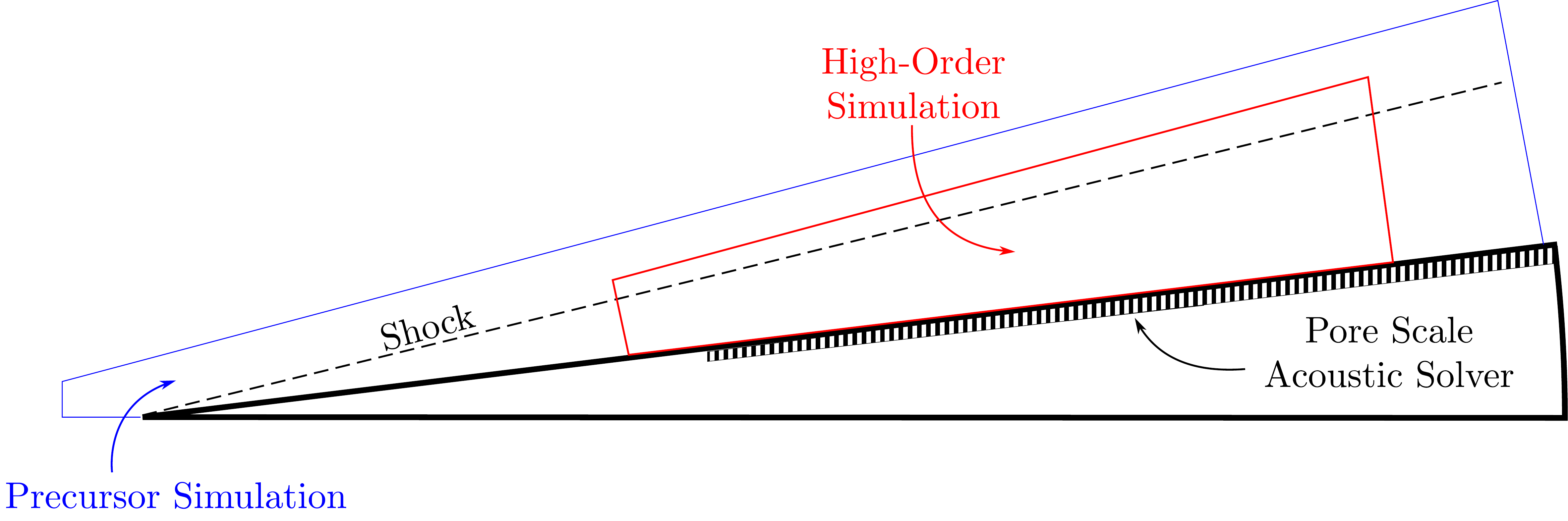}
\caption{Illustration of the three computational domains, corresponding to different numerical simulations carried out at different levels of fidelity: the flow computed by a steady, low-order precursor simulation (blue domain) is used to define the  inlet conditions for the high-order unsteady simulation (red domain), whose impedance boundary conditions at the wall modeling the C/C surfaces are informed by the aeroacoustic simulations carried out at a pore-scale level (striped domain) }
\end{center}
\vspace{-0.75cm}
\end{figure}

At the inflow, a Dirichlet boundary condition coupled with a buffer layer is used to input the flow profile calculated by the precursor and to prevent oscillations that could appear in the subsonic portion of the domain.
At the wall, Dirichlet conditions are used to impose no-slip, no penetration and suction and blowing when active.
Due to the small duration of the test run, the wall does not have time to heat up and thus can be assumed to be isothermal with a temperature of 300 K.
Neumann boundary conditions are imposed at the top and right boundaries for the velocity field.

A porous surface was introduced at the wall in a second campaign of runs via an impedance boundary condition (see equation \ref{eqn:impedance}).
The idea behind the implementation of a time domain IBC is to model the exact acoustic response of the UAC without the need to resolve its complex geometrical structure.This strategy allows for retaining high accuracy on the flow side by removing the  grid resolution requirements that would be needed to mesh the pores. 


\section{Computational Tools}

\subsection{Spectral-Element High-Order Unstructured Fully Compressible Solver: \underline{sd3DvisP}}

The \textsc{sd3DvisP} solver is an MPI parallelized fortran 90 code for compressible flows based on the high-order Spectral Difference (SD) scheme for unstructured hexahedral elements.~\cite{kopriva:96,sun:07}
The solver is capable of running with arbitrary preselected orders of accuracy and provides minimal numerical dissipation~\cite{lodato:14b,lodato:16,ChapelierLJ_CF_2016}.

The time integration is done explicitly with Runge-Kutta schemes which are up to 5th-order accurate.
Several different types of Runge-Kutta schemes are available in the present numerical solver, such as the standard second- and fourth-order schemes with two and four stages, respectively. 

A third-order three-stages TVD, the fourth-order five-stages SSP and the low-storage fourth-order six-stages low dispersion/dissipation RK schemes are also implemented.~\cite{gottlieb:98,spiteri:02,bogey:04}
The closures for LES modeling are provided by several eddy-viscosity models~\cite{smagorinsky:63,nicoud:99,nicoud:11,chapelier2016spectral} or by a similarity mixed model~\cite{lodato:09} coupled with dedicated constrained filtering operators of arbitrary order~\cite{lodato:13}. 
Wall modeling capabilities have also been designed for the SD setup.~\cite{lodato:14}
The code is also able to tackle high-speed compressible flows with shocks and discontinuities, using a self-calibrating shock capturing methodology that is available in the solver.
Originally designed for Discontinuous Galerkin schemes,~\cite{persson:06} this method enables sub-cell resolution of discontinuities detected via a modal sensor.
The \textsc{sd3DvisP} solver has been successfully applied to flows with shocks, including shock/vortex interaction and shock/wavy-wall interactions ~\cite{lodato:15,lodato:16} and nonlinear thermoacoustics \cite{GuptaLS_JFM_2017}.

\subsection{High-order structured compact-finite-difference solver in curvilinear coordinates: \underline{CFDSU}} \label{subsec:cfdsu}

CFDSU solves fully-compressible Navier-Stokes equations on a structured curvilinear grid using sixth-order compact and staggered finite difference scheme \cite{Lele_JCP_1992}.
Compact and staggered methods outperform conventional finite difference schemes at high wavenumbers, without restricting the geometry and boundary conditions of the problem demanded by spectral method \cite{jiang2009numerical,NagarajanLF_JCP_2003}. 

Based on the Navier-Stokes equations it is possible to write the governing equations of the flow in a curvilinear coordinate system $(x^1,x^2,x^3)$ based on the contravariant velocity terms $v^i = dx^i/dt$ but we also need to define the Cartesian coordinate system $(x_1,x_2,x_3)$ where the equation set is originally derived. The relations shown here were presented by Nagarajan et al \cite{nagarajan2007leading}. 

\begin{equation}
	\frac{\partial J \rho}{\partial t}  + \frac{\partial }{\partial x^j}( J \rho v^j)= 0
\end{equation}

\begin{equation}
	\frac{\partial J \rho v^i}{\partial t}  + \frac{\partial}{\partial x^j}(J \rho v^i v^j + J p g^{ij} - J \sigma^{ij})= 
	-J\Gamma^i_{qj}(\rho v^q v^j + p g^{qj} - \sigma^{qj})
\end{equation}

\begin{equation}
	\frac{\partial J E}{\partial t}  + \frac{\partial}{\partial x^j}(J(E + p) v^j + Jq^{j})= \frac{\partial}{\partial x^k}(J \sigma^{ij}g_{ik}v^k )
\end{equation}

In this frame of reference the total energy, the viscous stress tensor and the heat flux vector are described by slightly modified relations described below.

\begin{equation}
	E = \frac{p}{\gamma - 1} + \frac{1}{2} \rho g_{ij} v^i v^j
\end{equation}

\begin{equation}
	\sigma^{ij} = \mu (g^{jk} \frac{\partial v^i}{\partial x^k} + g^{ik} \frac{\partial v^j}{\partial x^k} - \frac{2}{3}g^{ij}\frac{\partial v^k}{\partial x^k})
\end{equation}

\begin{equation}
	q_i = -k  g^{ij} \frac{\partial T}{\partial x^i}
\end{equation}

In the notation used the superscripts indicate the contravariant tensors and the subscripts the covariant ones. Following the convention, $g^{ij}$ is the contravariant metric tensor and the covariant is $g_{ij}$. $\Gamma^i_{qj}$ is the Christoffel symbol of the second kind, which is the representation of a second order derivative of a vector or a first order derivative of a second order tensor and $J$ is the Jacobian of this transformation, which is the determinant of the Jacobi matrix ($J_{ij} = \partial x_i/\partial x^j$).

These metrics components are responsible for translating the curvilinear coordinates into its Cartesian component and accounting for the effects of using a non-inertial reference frame.

\begin{eqnarray}
	g^{ij} = \frac{\partial x_i \partial x_j}{\partial x^k \partial x^k},\  	
    g_{ij} = \frac{\partial x^k \partial x^k}{\partial x_i \partial x_j},\ 
    \Gamma^i_{qj} = \frac{\partial x_i}{\partial x^l}  \frac{\partial^2 x^l}{\partial x_q \partial x_j}.
\end{eqnarray}

An IBC was implemented in this high-order curvilinear code based on the derivations made by Fung and Yu \cite{fung2004time} and in the implementation on a fully compressible Navier-Stokes flow solver made by Scalo et al. \cite{ScaloBL_PoF_2015}. The specified conditions are implemented based on the relations between the incoming ($-$) and outgoing ($+$) waves which respective velocities read:

\begin{equation}
	\begin{cases}
   		 v^- = v' + \frac{p'}{\rho_0 a_0} \\
    	 v^+ = v' - \frac{p'}{\rho_0 a_0}
	\end{cases}
\end{equation}

For the correct implementation of the impedance boundary condition in curvilinear coordinates, $v'$ and $p'$ need to be the fluctuating values, in physical space, of wall normal velocity and pressure respectively. In addition, $\rho_0$ and $a_0$ are the base density and speed of sound of the fluid. 

Adopting a harmonic convention to transform the variables to frequency domain, the acoustic impedance $Z(\omega)$ is defined as a function of the perturbed pressure $p(t)$ and its induced normal velocity component $v(t)$ into a porous surface, satisfying the relation:

\begin{equation}
\hat{p}(\omega) = \rho_0 a_0 Z_*(\omega)\hat{v}(\omega),
\label{eqn:impedance}
\end{equation}

In the frequency domain, the outgoing wave can be related to the incoming one by the reflection coefficient, $\widehat{W}$ or the wall softness coefficient $\widehat{\widetilde{W}}$, defined as:

\begin{eqnarray}
	\widehat{W} = \frac{\rho_0 a_0 - Z_*(\omega)}{\rho_0 a_0 + Z_*(\omega)},\ 
     \widehat{\widetilde{W}} = \widehat{W}  + 1;
\end{eqnarray}

\begin{eqnarray}
	\hat{v}^+ = \widehat{W}\hat{v}^-,\ 
    \hat{v}^+ = - \hat{v}^- + \widehat{\widetilde{W}}\hat{v}^-.
    \label{eqn:outgoing}
\end{eqnarray}

It is interesting to note that a solid, or purely reflective, wall is obtained by setting $\widehat{\widetilde{W}} = 0$, which occurs at the limit $|Z| \rightarrow \infty$. With these relations defined, Fung and Ju \cite{fung2004time} pointed out that it is possible to solve the outgoing wave, based on equation \ref{eqn:outgoing}, by a causal convolution of the incident wave. As soon as the wave is solved, the wall normal velocity can be recovered and imposed as Dirichlet Boundary conditions at each time step. 

\begin{equation}
v' = \frac{1}{2}\left[ v^- + v^+\right]
\end{equation}

\subsection{Pore-Cavity Inverse Ultrasonic Solver: \underline{iHS}} \label{sec:tools:iuhs}

The inverse Helmholtz solver (iHS)~\cite{PatelEtAl_Aviation_2017,PatelEtAl_arxiv_2017} is a novel computational methodology that allows the evaluation of the spatial distribution of acoustic impedance at the open surface of an arbitrarily shaped cavity in response to acoustic wave propagation in the cavity at a given frequency. Multiple instances of the iHS can be concurrently executed to reconstruct the full, broadband acoustic impedance at the open surface of any given geometry. Such an impedance can then be implemented as a time-domain impedance boundary condition (IBC) in flow-side-only simulations as discussed in the previous section (\S \ref{subsec:cfdsu}).

\begin{figure}[htbp]
\centering
\includegraphics[width=0.3\linewidth]{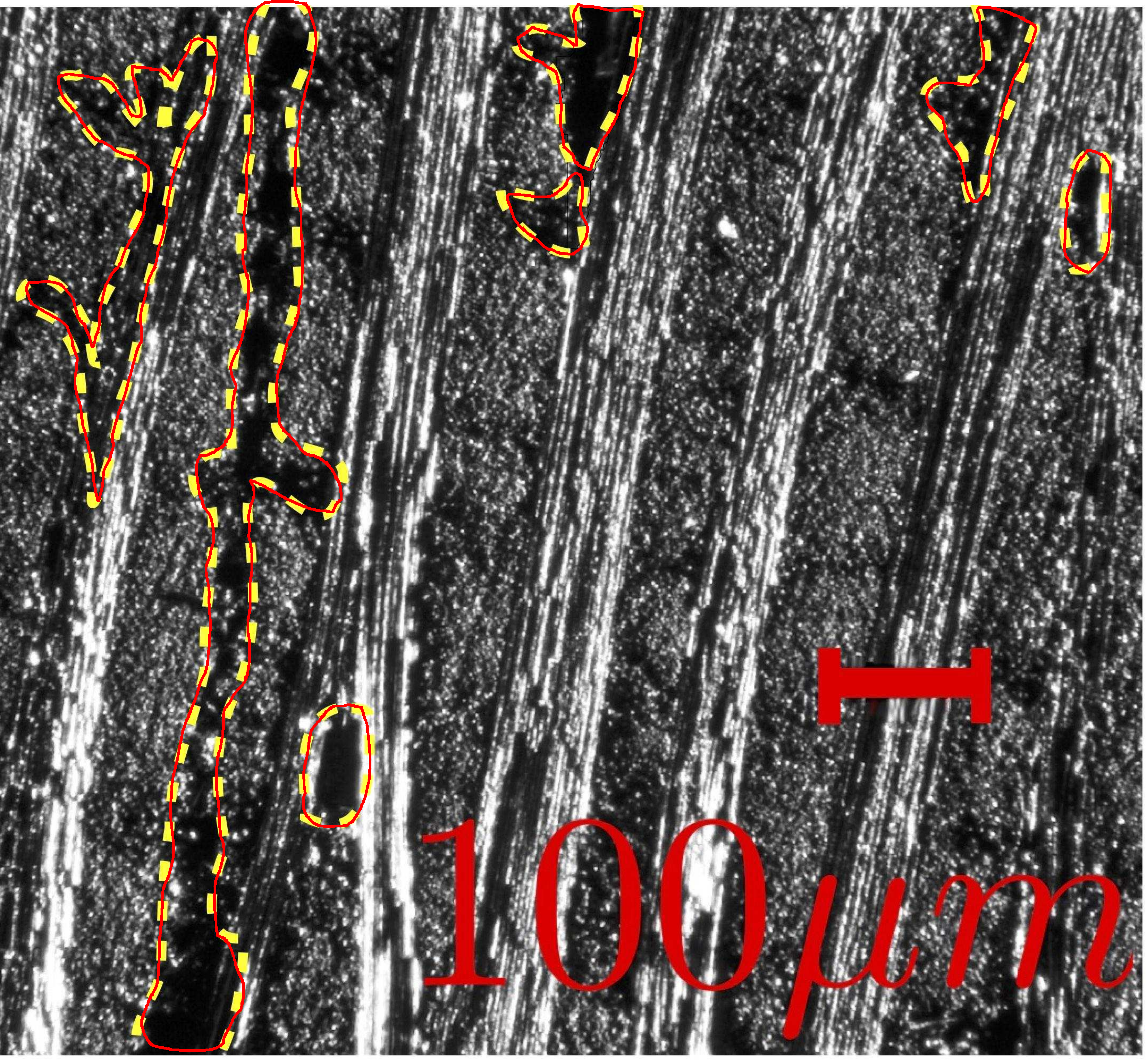}
\includegraphics[width=0.284\linewidth]{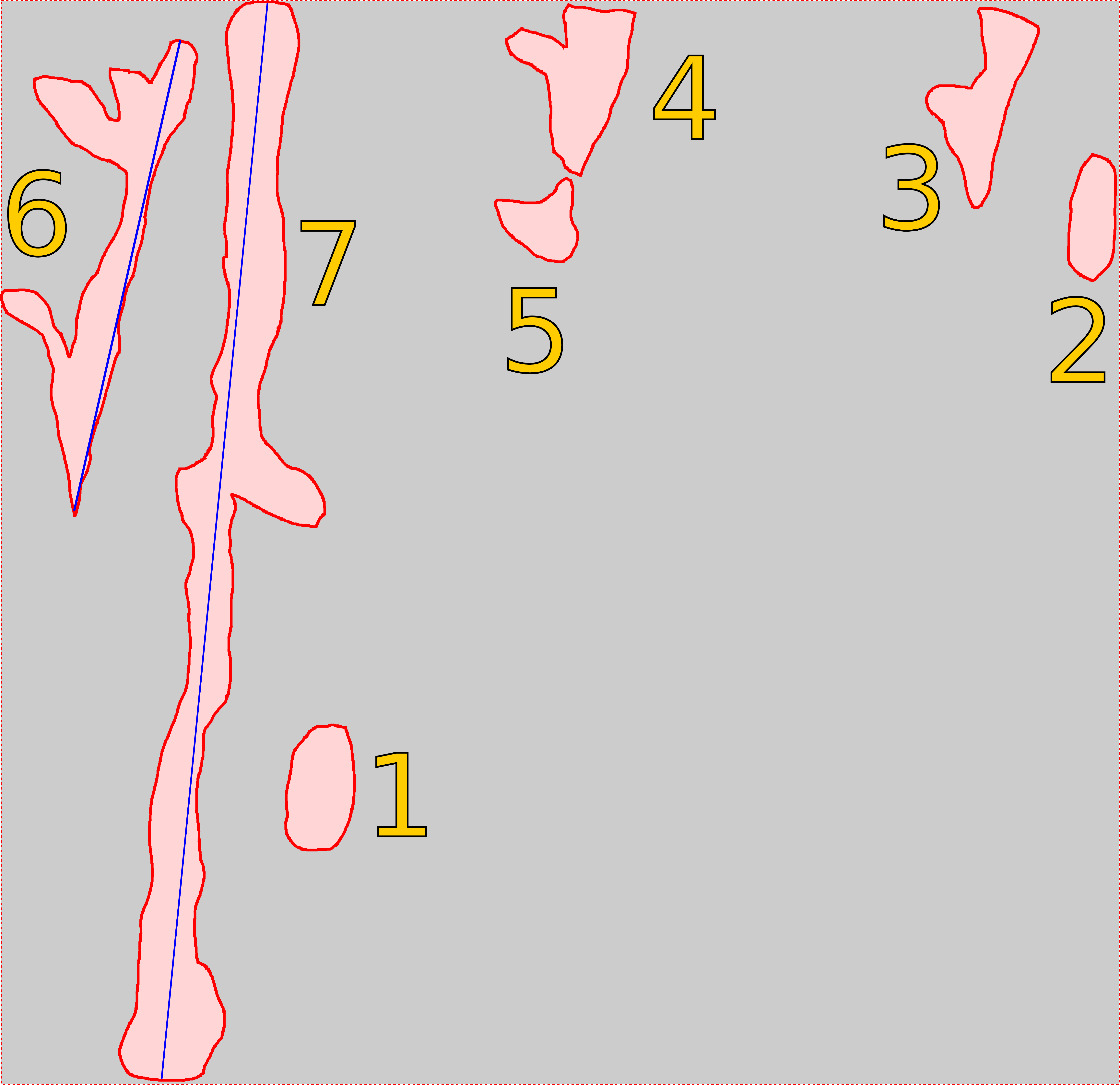}
\caption{Surface scan of a C/C sample manufactured at DLR Stuttgart (left); result of a semi-automatic edge-detection algorithm to identify the edges of the open cavities (right) responsible to ultrasonic wave absorption.}
\label{fig:cracks}
\end{figure}

From a high resolution image of the C/C block (figure 3), the edges of large surface cavities are auto-detected and their volumes meshed to be provided as an input geometry to the iHS, which evaluates the broadband impedance of each individual surface cavity.
These impedances are then combined, noting that the admittance of the surrounding hard surface is zero, to yield the effective surface averaged impedance for the C/C block in consideration.


\section{Results}

The goal of this section is to reproduce and analyze the disturbance amplification mechanism in a hypersonic flow over a sharp cone with solid walls through the excitation of the boundary layer with a broadband frequency pulse.
Then, we model the acoustic response of real C/C materials through the iHS technique explained in section III.C and test its attenuation properties in the presence of harmonic disturbances that excite the second mode transition mechanism inside the high speed boundary layer.

 \subsection{Broadband Pulse Disturbance Introduction and Advection over Solid Wall}

\begin{figure}[htbp]
\centering
\includegraphics[width=1.\linewidth]{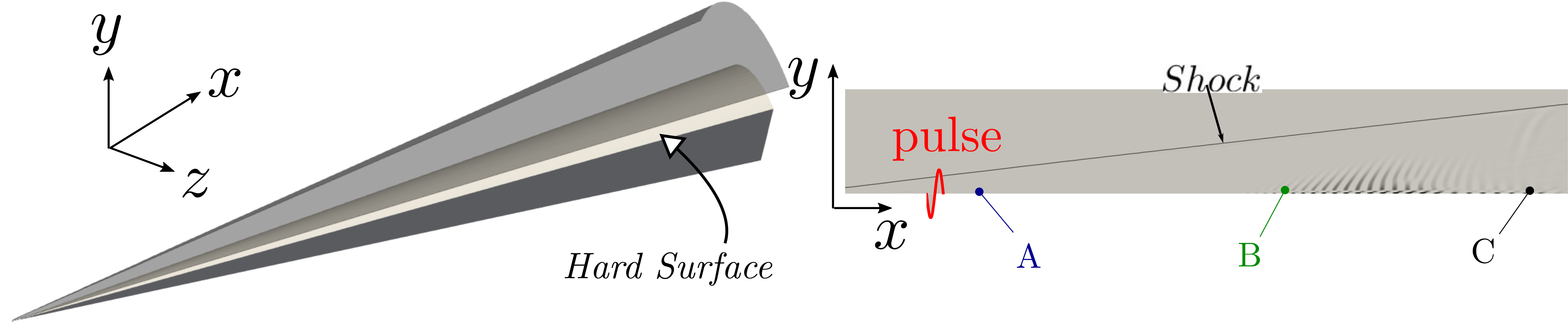}
\caption{Axi-symmetric computational setup for one-time broadband pulse excitation simulations over a 7$^{\textrm{o}}$ sharp cone at $M_\infty=7.4$ and $Re_m=4.06\,\times\,10^6 \left[ \textrm{1/m}\right]$. \label{fig:setup_pulse}} 
\end{figure}

A precursor computation of the cone is first performed using the solver sd3dvisP. The mesh consists of 330,000 non-regular hexahedral elements. The simulation is run at first order of accuracy, and the numerical fluxes at interfaces between elements are computed using the Roe flux with entropy fix, which introduces the right amount of numerical dissipation in the vicinity of shocks to stabilize the computation. Without perturbations, the flow establishes to a steady-state laminar hypersonic boundary layer.

After the establishment of the unperturbed steady base state, the data plane of the precursor computation is extracted to provide inlet boundary conditions to the high-order solver CFDSU. Two types of controlled disturbances are then imposed via suction and blowing at the bottom wall \cite{SivaFasel_JFM_2014}. The first one is a spatially and temporally broadband pulse applied for a finite amount of time (figure \ref{fig:setup_pulse}). This perturbation aims at mimicking a natural transition scenario and determining the most amplified frequency inside the boundary layer, and it was first used by Gaster and Grant \cite{gaster1975experimental} in incompressible boundary layer transition simulations and by Sivasubramanian et al.\cite{SivaFasel_JFM_2014} in hypersonics.  The pulse is applied for 3.3 $\mu s$ and the amplitude was chosen small enough to trigger only the linear growth mechanisms. 

Since not all the energy introduced by the pulse inside the computational domain is received by the boundary layer, we probe the signal near the pulse excitation location  and inside the boundary layer (probe A) to quantify the initial disturbance energy distribution. Two other probe locations inside the boundary layer are selected to identify the disturbance dynamics for this case.

\begin{figure}[htbp]
\centering
\includegraphics[width=1.\linewidth]{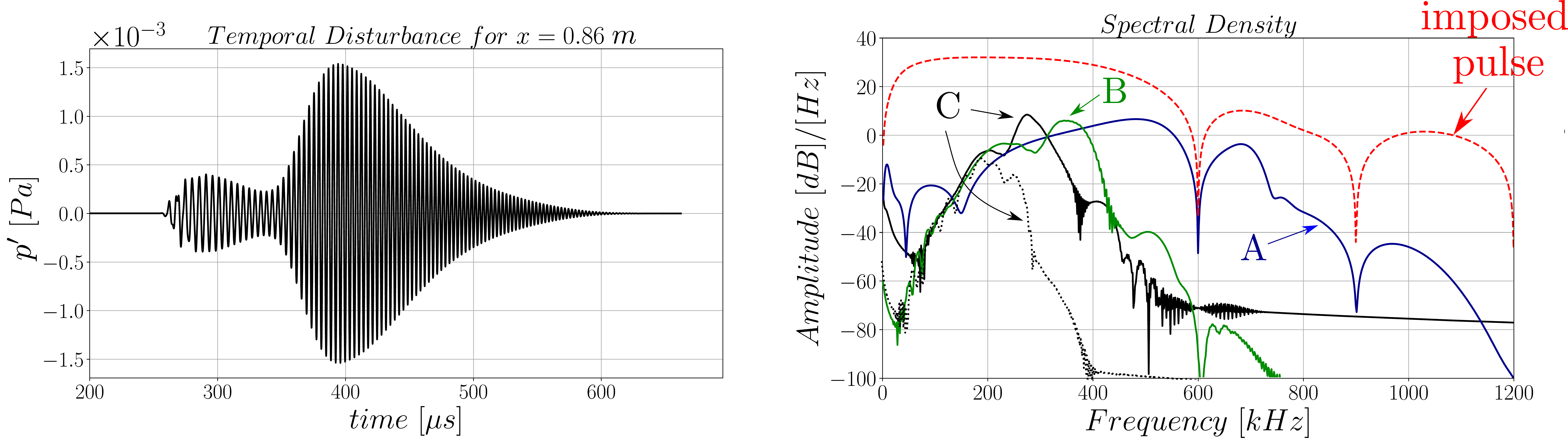}
\caption{Results of axisymmetric simulation of broadband pulse propagation over a 7$^{\textrm{o}}$ sharp cone at $M_\infty=7.4$ and $Re_m=4.06\,\times\,10^6 \left[ \textrm{1/m}\right]$. Results for probe C are shown for two different grid resolutions in the x direction. The dotted line ($:$) corresponds to a simulation with half of the resolution of the solid line ($-$), the selected grid.\label{fig:res_pulse}} 
\end{figure}
\vspace*{-0.7cm}
\begin{table}
\centering
\caption{Probe Locations and corresponding Frequency with Maximum Energy}

\begin{tabular}{lccc}
\hline \hline
 Probes & A & B  & C \\ 
\hline                               

$x \ [cm]$ & 19 & 68 & 86\\
$f \ [kHz]$ & 459 & 341 & 274\\

\hline \hline
\end{tabular}
\end{table}

The results shown in figure \ref{fig:res_pulse} demonstrate the narrowing of the energy spectrum after advection for both probes. This leads to the conclusion that a perturbation at a frequency not tuned to the boundary layer thickness at a particular location is rapidly damped. We can also observe that for probes B and C there is amplification of a certain band of frequencies in the ultrasonic regime characteristic of the second mode transition mechanism in hypersonic flows and consistent with results shown in Wagner's experiments \cite{Wagner:2014PhD}. 

The broadband pulse excitation performed is also able to capture and quantify the correlation between the most amplified frequencies in the boundary layer and its thickness. It is observed that, as the wave moves downstream into regions of higher thicknesses, the most amplified frequencies are shifted to lower values. This is consistent with the observation made by Stetson \cite{stetson1992example} that the second mode is highly tuned to the boundary layer thickness. 

The grid resolution used to run the aforementioned simulations was selected to allow the simulation of waves up to 450 kHz with a minimum amount of artificial diffusion. In figure \ref{fig:res_pulse}, we show that, although a run with a lower grid resolution is able to solve low frequency waves as well as the higher resolution run, it is not able to accurately simulate the advection of the ultrasonic waves corresponding to the second-mode transition mechanism.

\subsection{Impedance Characterization of Carbon-Fiber-Reinforced Carbon Ceramics (C/C) Materials}

\begin{figure}[htbp]
\centering
\includegraphics[width=0.25\linewidth]{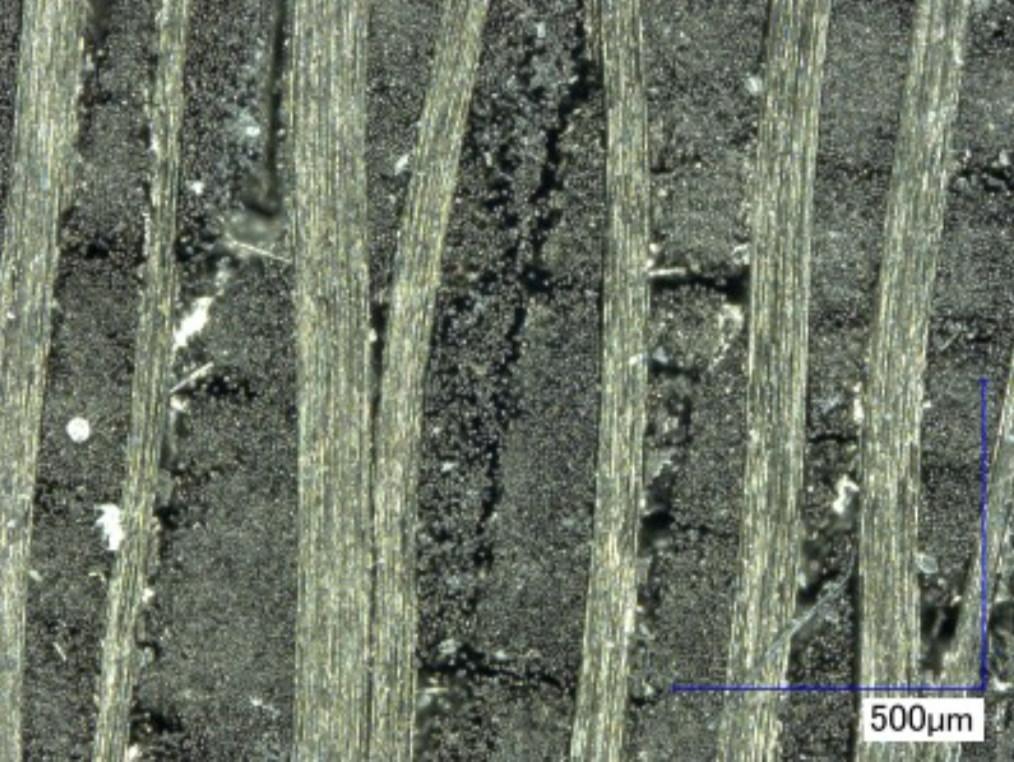}
\hspace{0.2cm}
\includegraphics[width=0.25\linewidth]{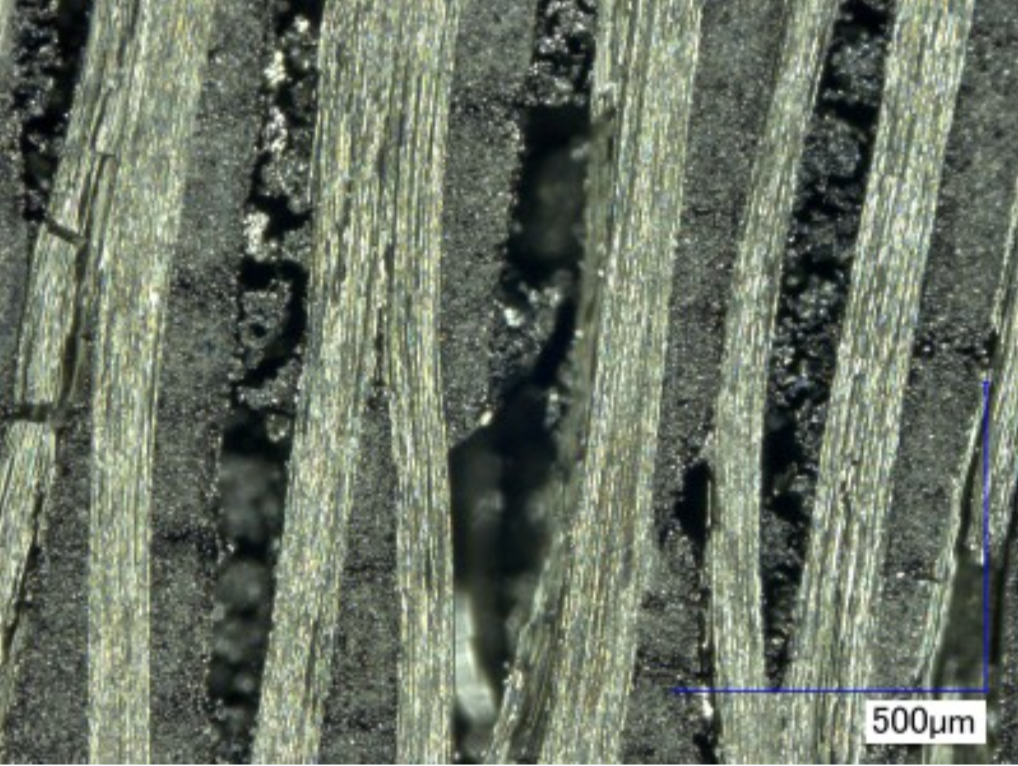}
\hspace{0.2cm}
\includegraphics[width=0.25\linewidth]{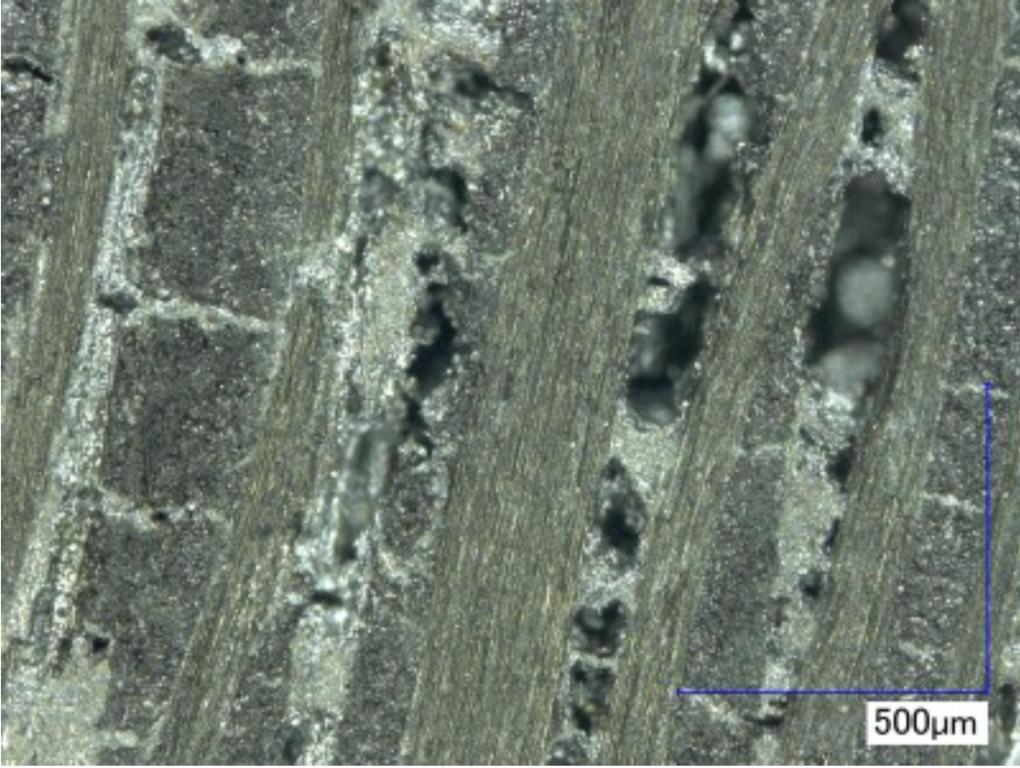}
\put(-322,-15){a)}
\put(-192,-15){b)}
\put(-60,-15){c)}
\caption{Samples of carbon fiber reinforced carbon ceramics materials\cite{wagner2015potential}. a) `Classical' C/C, b) Optimized C/C, c) C/C-SiC based in optimized C/C. \label{fig:cc_samples}} 
\end{figure}

C/C-SiC is a material that was already used as a Thermal Protection System (TPS) in a hypersonic flight \cite{TurnerHJST_AIAA_2006,WeihsLT_IAC_2008} and C/C represents an intermediate state of the its manufacturing process. The `classical' C/C porosity is a result of thermal stresses that appear in the cool down process of the material after the pyrolyzation of the matrix. This material offers excellent thermal resistance, low expansion and specific weight as well as high temperature stability at non oxidizing atmospheres. However, C/C can't be used when the oxidizing effects are important in the flow without a protective treatment. Resistance against oxidation is acquired by the infiltration of a liquid phase of silicon into the porous carbon, followed by a reaction to SiC. If this infiltration is done in the `classical' C/C, its microstructural gaps are mostly closed and the acoustic absorption properties of the material is greatly reduced. 

To solve this problem, DLR Stuttgart developed a technique to insert cavities into the carbon fiber fabrics during the processing of C/C which allows a doubling of the raw material porosity. The pre-existence of larger gaps in this optimized C/C allows it to retain a certain porosity degree after the SiC reaction. Overall, the optimized C/C-SiC combines the properties of oxidation resistance with acoustic absorption and, therefore, it is a suitable material for in flight transition delay.

\begin{figure}[htbp]
\begin{center}
\includegraphics[width =0.64\columnwidth]{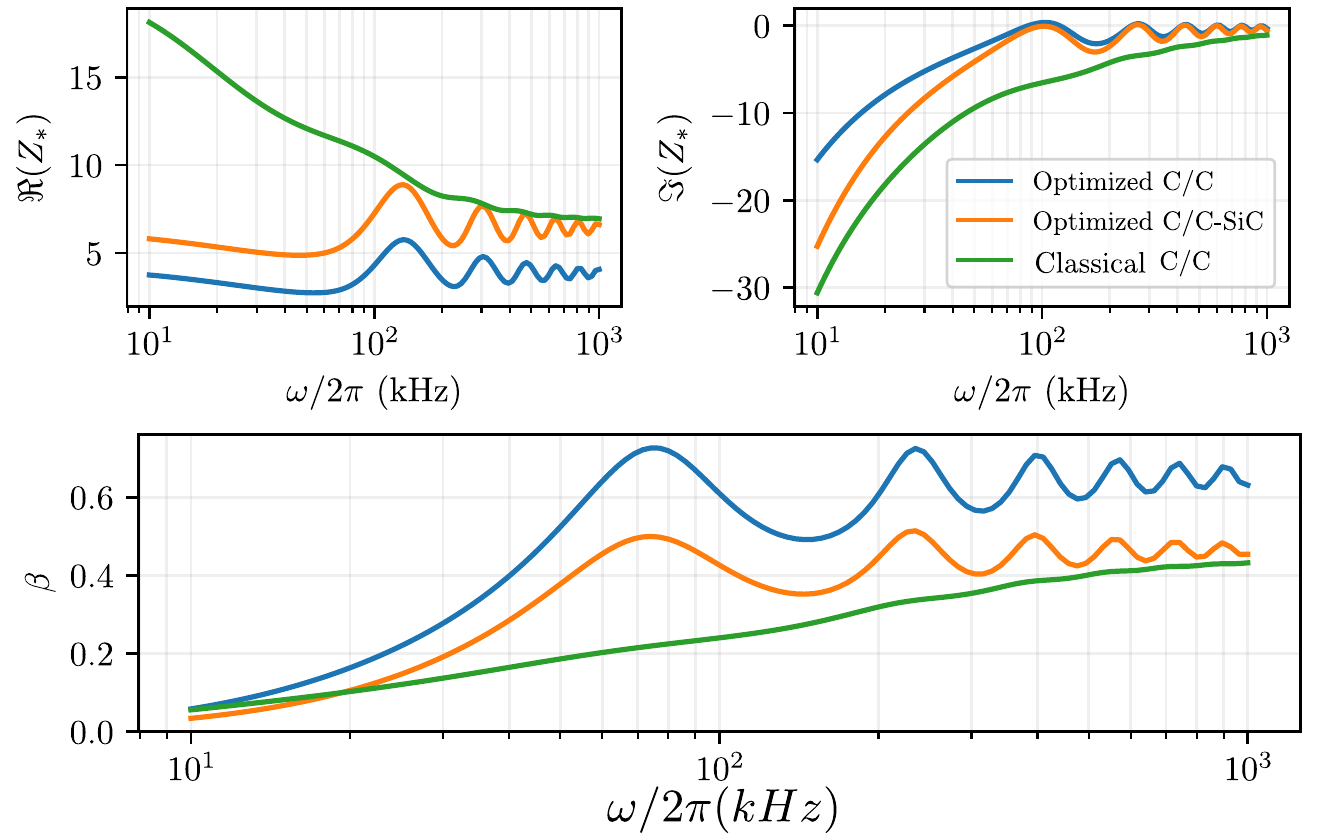}
\caption{
Real $\Re{(Z_*)}$ and imaginary $\Im{(Z_*)}$ components of the complex specific impedance $Z_*(\omega)$ calculated using the iHS technique (see section \ref{sec:tools:iuhs}) for the three different C/C-based materials shown in figure \ref{fig:cc_samples}. Absorption coefficient, $\beta$ \eqref{eq:absorption_coefficient}  is plotted below versus frequency. }
\label{fig:beta}
\end{center}
\end{figure}

The acoustic impedance of the samples shown in figure \ref{fig:cc_samples} were approximated via the iHS methodology (section III.C) and are shown, in figure \ref{fig:beta}, to have a real and imaginary component. Therefore, it is expected that a wave interacting with a porous surface made from these materials will not only have its magnitude decreased but will also experience a phase change. With the characterization of the acoustic impedance at the mouth of the cavities, we are also able to reconstruct the absorption coefficient ($\beta$),

\begin{equation} \label{eq:absorption_coefficient}
\beta(\omega) = 1 -  \left|\frac{1-Z_*(\omega)}{1+Z_*(\omega)} \right|^2.
\end{equation}

$\beta$ is defined as the fraction of incident energy absorbed at the surface and we can see that the higher the material porosity, the higher its capacity to absorb energy. Other observable trends are that the presence of larger porous cavities in the optimized C/C and optimized C/C-SiC lead to a resonance-like behavior and that higher frequency strengthens the thermoviscous dissipation effects. 

These results were obtained for the pressure conditions at the surface of our conical model, i.e. the conditions after the shock, which were of approximately 5000 Pa. These conditions operate in a low pressure range, which is not fully covered by the bench experiments performed by Wagner et al. \cite{Wagner2014}. However, the qualitative trend of the optimized C/C having the highest absorption capability amongst the three materials and the `classical' C/C having the lowest is observable in both experimental and numerical data.

\subsection{Harmonic Disturbance Excitation over Solid and Porous Walls}

Once the unstable frequency band and its spatial variation is known, a harmonic oscillator (second type of disturbance) is used to drive tonal disturbances into the boundary layer with a spatial distribution \cite{harris1997numerical} meant to steadily excite only the desired frequency while allowing for a natural convective growth. Two frequency levels, $f = 270$ kHz and $f = 350$ kHz, were chosen to highlight the transition mechanism dependency on this parameter. 

\begin{figure}[htbp]
\centering
\includegraphics[width=1.\linewidth]{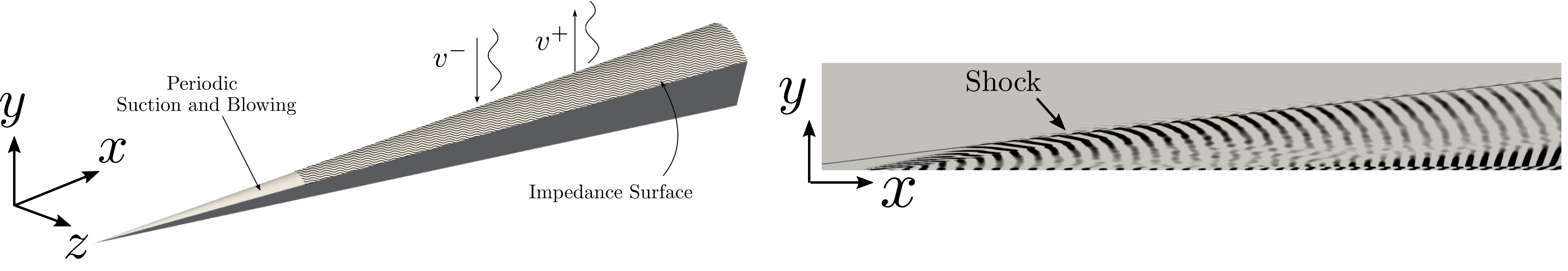}
\caption{Computational setup for harmonically forced hypersonic transition simulation over impedance (left) and contours of the dilatational field (right) from a simulation over for impermeable (or solid) walls on a 7$^{\textrm{o}}$ sharp cone at $Re_m=4.06\,\times\,10^6 \left[ \textrm{1/m}\right]$.}\label{fig:harmonic_intro} 
\end{figure}

\begin{figure}[htbp]
\centering
\includegraphics[width=1.\linewidth]{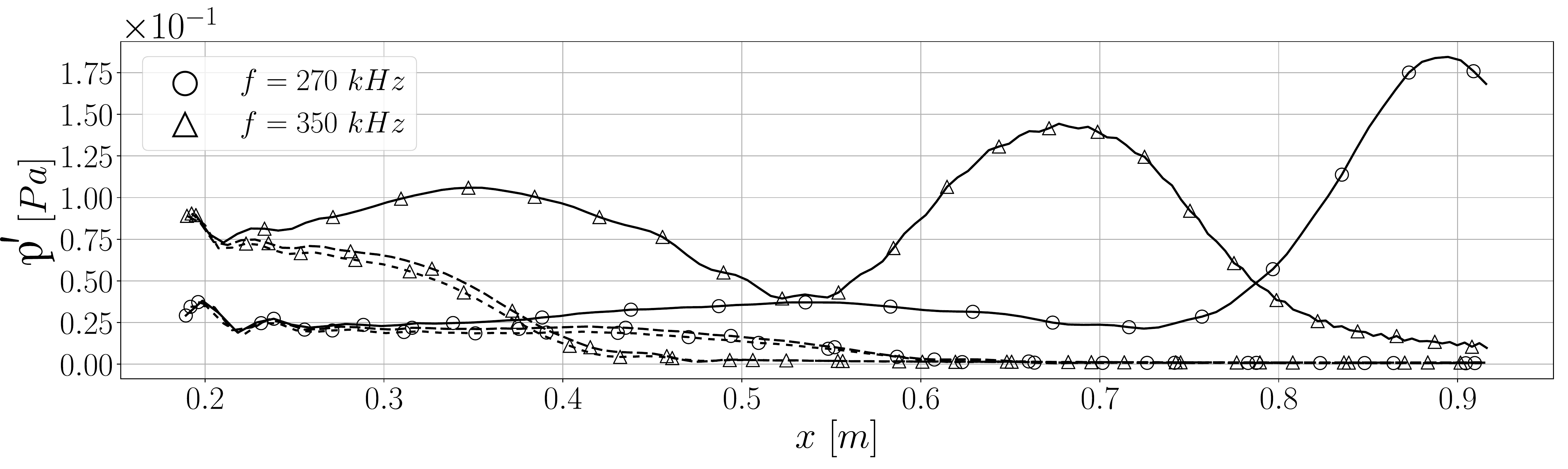}
\caption{Spatial distribution of pressure disturbance amplitude at the wall for harmonically forced simulations at two different frequencies. Results are shown for solid ($-$) and porous walls, including `Classical' C/C  ($---$) and C/C-SiC ($-\ -\ -$) . }\label{fig:harmonic} 
\end{figure}

All the harmonic excitation cases were performed with the same small amplitude input and spatial distribution meant to excite only linear mechanisms with the goal of studying only the effects of frequency and surface porosity into the artificially introduced perturbation. Even though the same amount of energy was inserted in the boundary layer for both frequencies, we observe in figure \ref{fig:harmonic} that, at the same distance downstream of the excitation location, the $f = 350$ kHz case has a higher amplitude of pressure oscillation than $f = 270$ kHz. This is an evidence of variations in receptivity of the boundary layer to external perturbations of different frequencies.

An analysis of the region downstream of the harmonic excitation location for the two different frequencies studied reveals that the results of disturbance advection over solid surfaces are consistent with the broadband pulse test previously performed. Figure \ref{fig:harmonic} shows that the amplification region for the higher frequency mode is upstream of the lower frequency, that can also be observed in figure \ref{fig:res_pulse}. 

In comparison with the spatial growth of the uncontrolled cases, the presence of a porous surface, modeled as an impedance boundary condition (IBC) that mimics the acoustic influence of real C/C samples in the flow, was capable damping completely the second mode-energy (figure \ref{fig:harmonic}) for all cases presented. It is interesting to notice that, even though the high frequency case has a higher initial amplitude due to a receptivity difference, this mode was damped spatially faster than the low frequency mode. This result is consistent with the trend presented in figure \ref{fig:beta}, that shows that the absorption coefficient ($\beta$) increases with frequency.

As a final remark, for the frequencies and flow condition tested, the influence of the `classical' C/C and the optimized C/C-SiC in the flow are similar but figure \ref{fig:harmonic} shows that the higher porosity and $\beta$ of the siliconized carbon-carbon makes it capable of stabilizing the flow slightly faster.

\section{Summary}

The effect of realistic acoustically absorptive surfaces on the second mode in a hypersonic boundary layer over a sharp slender cone was numerically studied in the present work. A combination of precursor simulations and high order simulations were performed. The precursor was used to obtain good initial conditions to the more accurate and better resolved simulation. In addition, a novel technique (iHS) was used to solve the acoustic response at the mouth of the cavities present in porous materials and to model their collective influence through a surface averaging. The complex impedance obtained as a result was used to model the impact of the ultrasonically absorptive surface in a spatially developing boundary layer simulation without having to resolve the complex porous structure. 

Artificial disturbances were introduced in the boundary layer and their evolution with downstream advection was observed. With the introduction of a broadband pulse and its advection over a solid surface we were able to identify the second-mode unstable frequency band for the case studied. This information was used to drive harmonic excitation simulations over a solid wall and over materials currently being developed to be used in hypersonic transition control, the `classical' C/C and the optimized C/C-SiC.

In conclusion, for the frequencies and flow condition tested, both `classical' C/C and the optimized C/C-SiC were found capable of attenuating the second-mode and eventually stabilizing the boundary layer. These results, along with the fact that 'classical' C/C was already proven experimentally to delay transition\cite{Wagner2013a}, makes the optimized C/C-SiC a promising material for in flight transition control.

\section*{Acknowledgments}

We acknowledge the support of the Rosen Center for Advanced Computing (RCAC) at Purdue, the Air Force Office of Scientific Research (AFOSR) grant FA9550-16-1-0209, the AFOSR YIP 2018 and the very fruitful discussions with Dr. Ivett Leyva (AFOSR). Victor Sousa also acknowledges the support of the prestigious Lynn Fellowship administered by the interdisciplinary Computational Science and Engineering (CS\&E) graduate program at Purdue University.



\bibliographystyle{aiaa}


\end{document}